\documentclass[aps,prl,superscriptaddress,twocolumn]{revtex4-1}

\usepackage{graphics}
\usepackage{graphicx} 
\usepackage{dcolumn}
\usepackage{bm}
\usepackage{epstopdf}

\begin{document} 
	

\title{Strong anisotropy within a Heisenberg model in the ${J}_{\mathrm{eff}}\mathbf{=}1/2$ insulating state of Sr$_2$Ir$_{0.8}$Ru$_{0.2}$O$_4$}



\author{S.~Calder}
\email{caldersa@ornl.gov}
\affiliation{Quantum Condensed Matter Division, Oak Ridge National Laboratory, Oak Ridge, TN 37831.}

\author{J.~W.~Kim}
\affiliation{Advanced Photon Source, Argonne National Laboratory, Argonne, IL 60439.}

\author{A.~E.~Taylor}
\affiliation{Quantum Condensed Matter Division, Oak Ridge National Laboratory, Oak Ridge, TN 37831.}

\author{M. H. Upton}
\affiliation{Advanced Photon Source, Argonne National Laboratory, Argonne, Illinois 60439, USA}

\author{D. Casa}
\affiliation{Advanced Photon Source, Argonne National Laboratory, Argonne, Illinois 60439, USA}

\author{Guixin Cao}
\affiliation{Department of Materials Science and Engineering, University of Tennessee, Knoxville, TN 37996.}
\affiliation{Materials Science and Technology Division, Oak Ridge National Laboratory, Oak Ridge, TN 37831.}

\author{M.~D.~Lumsden}
\affiliation{Quantum Condensed Matter Division, Oak Ridge National Laboratory, Oak Ridge, TN 37831.}
 
\author{A.~D.~Christianson}
\affiliation{Quantum Condensed Matter Division, Oak Ridge National Laboratory, Oak Ridge, TN 37831.}

 

\begin{abstract}
The dispersive magnetic excitations in Sr$_2$IrO$_4$ have previously been well described within an isospin-1/2 Heisenberg model on a square lattice that revealed parallels with La$_2$CuO$_4$. Here we investigate the inelastic spectra of Sr$_2$Ir$_{0.8}$Ru$_{0.2}$O$_4$ with resonant inelastic x-ray scattering (RIXS) at the Ir L$_3$-edge. The results are  well described using linear spin-wave theory within a similar Heisenberg model applicable to Sr$_2$IrO$_4$, however the disorder induced by the substitution of 20$\%$ Ir$^{4+}$ ions for Ru$^{4+}$ removes longer range exchange interactions. A large spin-gap (40 meV) is measured indicating strong anisotropy from spin-orbit coupling that is manifest due to the altered magnetic structure in Sr$_2$Ir$_{0.8}$Ru$_{0.2}$O$_4$ with c-axis aligned moments compared to the basal plane moments in the parent. Collectively the results indicate the robustness of a Heisenberg model description even when the magnetic structure is altered and the ${J}_{\mathrm{eff}}\mathbf{=}1/2$ moments diluted.
 \end{abstract}
 

\maketitle

Understanding the underlying mechanisms that create unconventional superconductivity remains one of central unsolved problems in condensed matter physics and has driven research on cuprates for decades. The search for related materials with similar phenomena in which to either observe superconductivity, or explain the lack thereof, has recently led to the investigation of Sr$_2$IrO$_4$ \cite{KimScience,PhysRevLett.108.177003,PhysRevLett.106.136402}. Concurrently iridates, and $5d$-based transition metal oxides in general, have undergone increased interest unrelated to superconductivity where the delicate balance of competing interactions of strong spin-orbit coupling (SOC), increased hybridization and reduced Coulomb interactions lead to novel magnetic phenomena \cite{annurev-conmatphys-020911-125138}. 

Sr$_2$IrO$_4$ hosts a spin-orbit entangled ground state with ${J}_{\mathrm{eff}}\mathbf{=}1/2$ magnetic moments \cite{KimScience}. Mapping of these pseudo-spins onto several distinct crystal structures has led to the uncovering of a variety of novel properties, such as Kitaev physics and Weyl semimetals \cite{PhysRevLett.102.017205,ChunNatPhys,ArnabRuCl3,PhysRevB.83.205101,PhysRevLett.117.037201}. Specific interest related to Sr$_2$IrO$_4$ has been the predictions of unconventional superconductivity driven by analogous  structural and spin properties found in the cuprates. In particular, proximity to a superconducting phase was strongly suggested by the  magnetic excitation spectra which revealed striking similarities to La$_2$CuO$_4$ and was able to be described within a pure Heisenberg model consisting of spin-1/2 on a square lattice \cite{PhysRevLett.86.5377,PhysRevLett.108.177003}. A further surprising aspect of the excitation spectra for Sr$_2$IrO$_4$ was the absence of a spin-gap. This would be expected to be large in a SOC dominated system with strong exchange interactions and was explained as being due to an accidental balance of symmetric and antisymmetric anisotropic terms for the basal plane moment case \cite{PhysRevB.89.180401}.

Parallels of Sr$_2$IrO$_4$ with the cuprates are continuing to mount as was shown with the observation of fermi arcs in surface doped Sr$_2$IrO$_4$ \cite{Kimfermiarcs}. Given the various similarities would suggest bulk superconductivity in the iridates is achievable. Although similarly a consistent lack of superconductivity would also be revealing in finding the key ingredients required for high T$\rm _c$ superconductivity amongst the plethora of candidate phenomena. As with cuprates to achieve bulk superconductivity in iridates a promising route is to dope the parent compound \cite{PhysRevLett.106.136402}. Moreover in the case of Sr$_2$IrO$_4$ doping has proven instructive in understanding the ${J}_{\mathrm{eff}}\mathbf{=}1/2$ moments as the magnetic interactions are altered and diluted. For example studies of the series Sr$_2$Ir$_{\rm 1-x}$Ru$\rm _x$O$_4$ and Sr$_2$Ir$_{\rm 1-x}$Rh$\rm _x$O$_4$ have found a percolative description appropriate as the Ir ions are diluted in explaining the presence of magnetism up to relatively high doping of 20-30$\%$  \cite{PhysRevB.89.054409,PhysRevB.92.165128}. Moreover, with regards to the explanation of the spin-gap in Sr$_2$IrO$_4$ perturbing the system to break the purported accidental degeneracy that suppresses the spin-gap will allow this reasoning to be directly tested.

Here we investigate spin excitations as the magnetic interactions in Sr$_2$IrO$_4$ are tuned by substitution of  Ru$^{4+}$ onto the Ir$^{4+}$ site in the form Sr$_2$Ir$_{0.8}$Ru$_{0.2}$O$_4$. This region of the phase diagram has well defined ${J}_{\mathrm{eff}}\mathbf{=}1/2$ magnetic moments along the $c$-axis, as opposed to the basal plane in undoped Sr$_2$IrO$_4$ \cite{PhysRevB.92.165128}, and is an insulator within the relativistic  ${J}_{\mathrm{eff}}\mathbf{=}1/2$ Mott-like description of the parent \cite{KimScience, PhysRevB.92.165128}. The results presented here reveal the presence of robust but damped spin-waves in this doped system that are well described within a Heisenberg model with a relative renormalization of the measured dispersion with some qualitatively similar features towards that observed for La$_2$CuO$_4$. Distinct from Sr$_2$IrO$_4$ and La$_2$CuO$_4$, however, is the observation of a large spin-gap indicating strong anisotropy. 

\begin{figure}[tb]
	\centering                      
	\includegraphics[trim=0.2cm 2.3cm 0.0cm 8.1cm,clip=true, width=0.9\columnwidth]{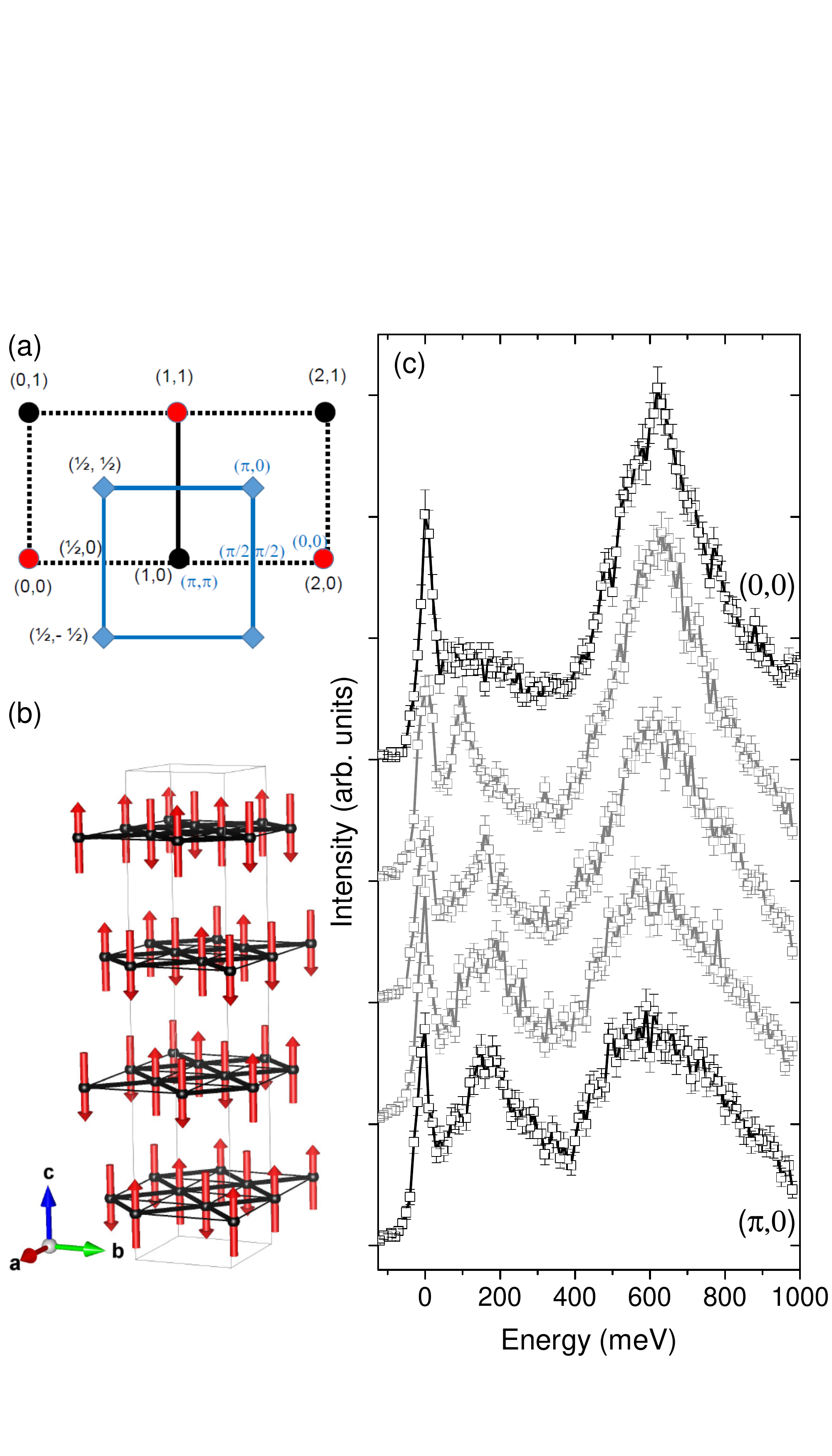}
	\caption{\label{Fig9ID}  (a) Structural (red circles) and magnetic (black circles) reciprocal space. High symmetry magnetic zone boundary points are indicated by the blue diamonds. H and K points are labeled along with the labeling in the square lattice notation used to describe the dispersions. The square lattice is rotated from the conventional lattice of Sr$_2$Ir$_{0.8}$Ru$_{0.2}$O$_4$. (b) Magnetic ground state of Sr$_2$Ir$_{0.8}$Ru$_{0.2}$O$_4$ has spins (red arrows) aligned along the $c$-axis. The nearest and next nearest bonds are shown as thick and thin black lines, respectively. (c) RIXS measurements at the Ir L-edge along several high symmetry directions in the low resolution set-up ($\Delta$E=90 meV).}
\end{figure}

To access the excitation spectra resonant inelastic x-ray scattering (RIXS) measurements at the Ir L$_3$-edge (11.215 keV) were performed at the Advanced Photon Source (APS), Argonne National Laboratory.  To aid comparison with separate studies on doped and undoped Sr$_2$IrO$_4$ and La$_2$CuO$_4$ we use the tetragonal $I4/mmm$ space group to describe reciprocal space. A map of the structural and magnetic reciprocal space is shown in Fig.~\ref{Fig9ID}(a). The magnetic structure of Sr$_2$Ir$_{0.8}$Ru$_{0.2}$O$_4$ has been well characterized using both polarized and unpolarized neutron diffraction and resonant x-ray scattering to consist of $c$-axis aligned spins \cite{PhysRevB.92.165128}. The magnetic structure is shown in Fig.~\ref{Fig9ID}(b). 

\begin{figure}[tb]
	\centering                      
	\includegraphics[trim=0.0cm 0.0cm 0.0cm 0.0cm,clip=true, width=0.9\columnwidth]{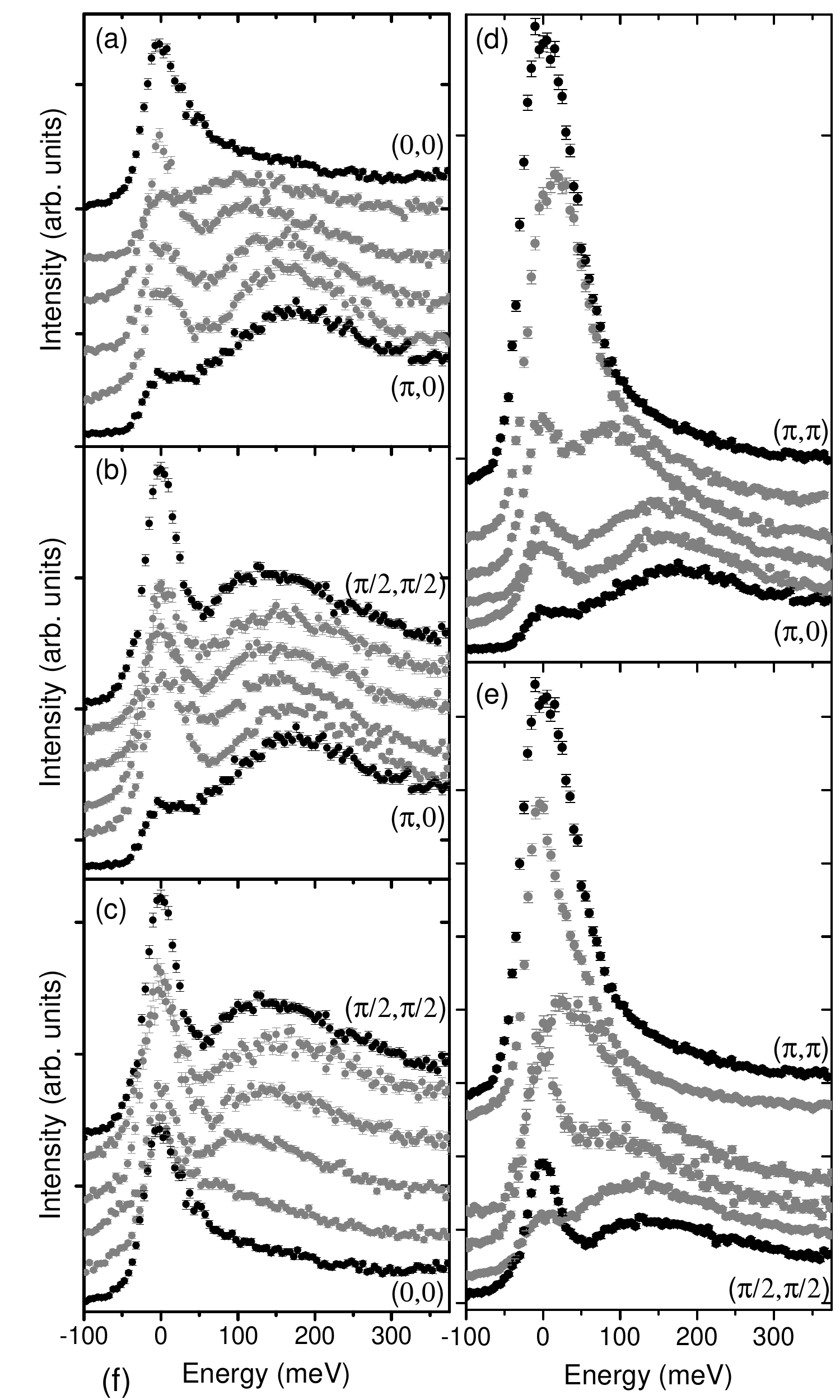}
	\includegraphics[trim=0.0cm 0.0cm 1.5cm 0.0cm,clip=true, width=0.75\columnwidth]{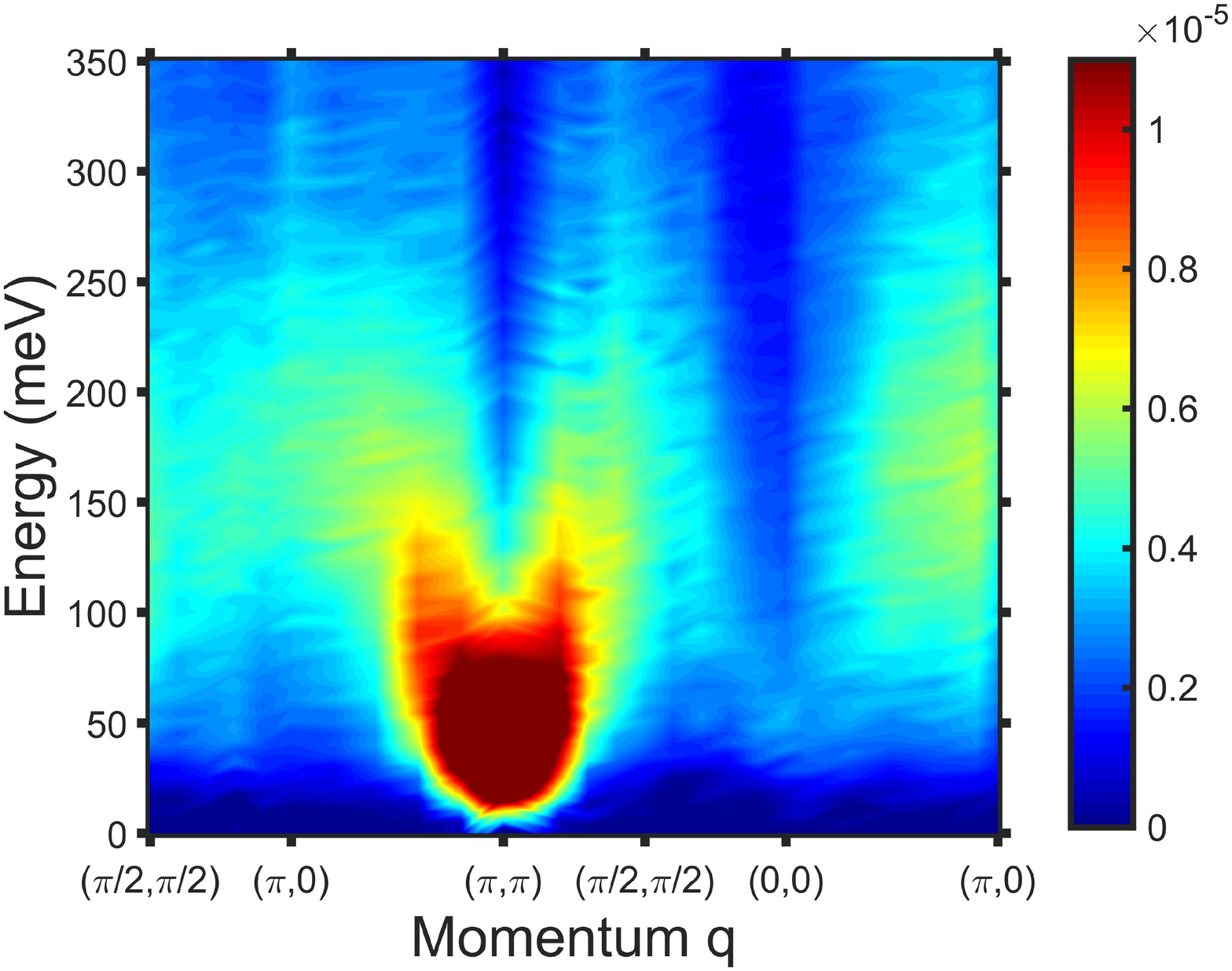}
	\caption{\label{Figscans} (a)-(e) High resolution RIXS measurements ($\Delta$E=35 meV) at the Ir L$_3$-edge following the low energy magnetic excitation along several high symmetry directions. (f) RIXS map of the low energy spectra with the elastic line subtracted.}
\end{figure}

Preliminary RIXS measurements were performed on sector 9-ID, followed by detailed higher energy resolution measurements with $\Delta$E=35 meV FWHM, based on fitting the quasi-elastic line to a charge peak, on MERIX at the APS \cite{MERIX}.  The scattering plane and incident photon polarization were both horizontal, i.e.~$\pi$ incident polarization, with the incident beam focused to a size of 41 $\times$ 25 $\mu$m$^2$ (H$\times$V) at the sample position on MERIX. To minimize elastic scattering measurements were performed with 2$\theta$ close to 90 degrees. This was achieved throughout the Brillouin zone by assuming L independence, in-line with previous RIXS measurements performed \cite{PhysRevLett.108.177003,PhysRevB.93.241102,PhysRevLett.117.107001,Ruarxiv}. The single crystal of Sr$_2$Ir$_{0.8}$Ru$_{0.2}$O$_4$ was grown and characterized as described in Ref. \onlinecite{PhysRevB.92.165128}. The initial 9-ID measurements,  with a resolution of $\Delta$E=90 meV FWHM, were used to follow higher energy excitations involving $d$-$d$ transitions. RIXS measurements to 1 eV energy loss are shown in Fig.~\ref{Fig9ID}(c). An inelastic excitation is observed around 650 meV. This excitation has been well characterized in the undoped compound as a spin-orbit exciton arising from transitions between the ${J}_{\mathrm{eff}}$ ground states and the presence here confirms that the ${J}_{\mathrm{eff}}\mathbf{=}1/2$ model remains unbroken in Sr$_2$Ir$_{0.8}$Ru$_{0.2}$O$_4$ \cite{KimScience}. Additionally the intensity variation, with a maximum at (0,0), corresponds to the behavior of Sr$_2$IrO$_4$. As with Sr$_2$IrO$_4$ a lower energy excitation is also observed below 200 meV. This appears to show some dispersion but given the resolution the results are hard to accurately interpret, therefore this scattering was probed with higher resolution RIXS.

To follow the dispersions below 400 meV RIXS measurements with the high resolution set-up ($\Delta$E=35 meV) were performed along high symmetry directions in the magnetic Brillouin zone as shown in Fig.~\ref{Figscans}(a)-(e). The excitations are strongly damped, particularly towards the zone boundary, as would be expected for a doped system. A dispersive magnetic excitation was observed with maxima at zone boundary and minima at zone center typical of a spin-wave excitation. Fitting the elastic line to a Lorentzian with width restricted to the instrumental resolution   and the excitation at zone center ($\pi,\pi$)  revealed a large spin gap of 40 meV. This can be seen qualitatively in the RIXS map that shows the scattering with the elastic line removed, Fig.~\ref{Figscans}(f).

The measured dispersion for Sr$_2$Ir$_{0.8}$Ru$_{0.2}$O$_4$ shows deviations from Sr$_2$IrO$_4$ in several key regards. (i) the ($\pi,0$) zone boundary maximum is reduced by $20\%$ to 156 meV indicating a reduction of exchange interactions. (ii) The ($\pi,0$) $\rightarrow$  ($\pi/2,\pi/2$)  dispersion softens since the ($\pi/2,\pi/2$) value is close to the parent but the  ($\pi,0$) is appreciably reduced. This has implications on exchange interactions beyond nearest and next nearest. (iii) The ($0,0$) $\rightarrow$  ($\pi,\pi$) nodal direction is hardened and the ($0,0$) $\rightarrow$  ($\pi,0$) antinodal direction  is softened compared to Sr$_2$IrO$_4$, directly opposite to the  behavior of electron doping on the Sr site \cite{PhysRevB.93.241102,PhysRevLett.117.107001}. (iv) The excitation is strongly gapped.

\begin{figure}[tb]
	\centering                      
	\includegraphics[trim=0.4cm 0.6cm 0.6cm 1.7cm,clip=true, width=0.9\columnwidth]{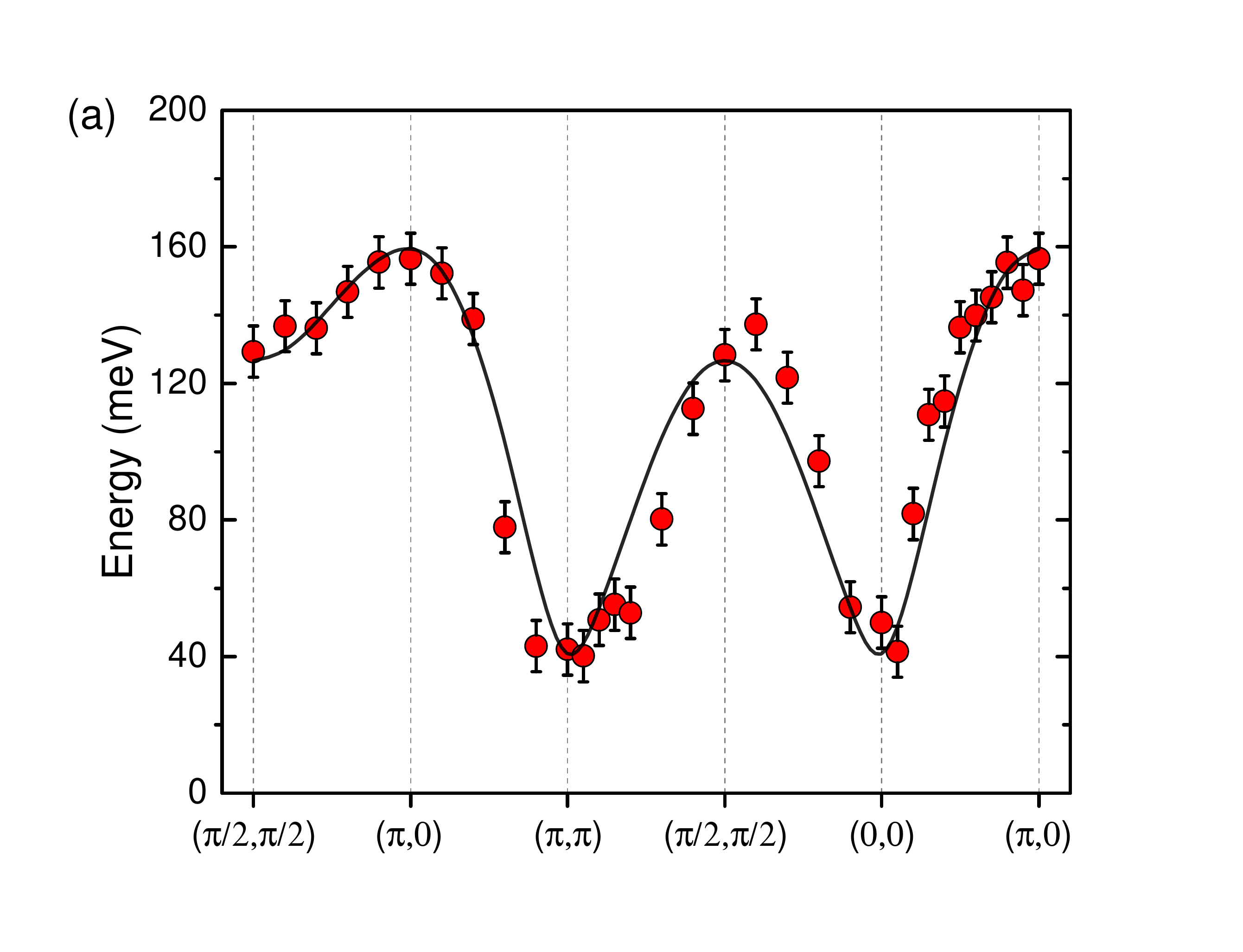}
	\includegraphics[trim=0.4cm 0.6cm 0.6cm 1.7cm,clip=true, width=0.9\columnwidth]{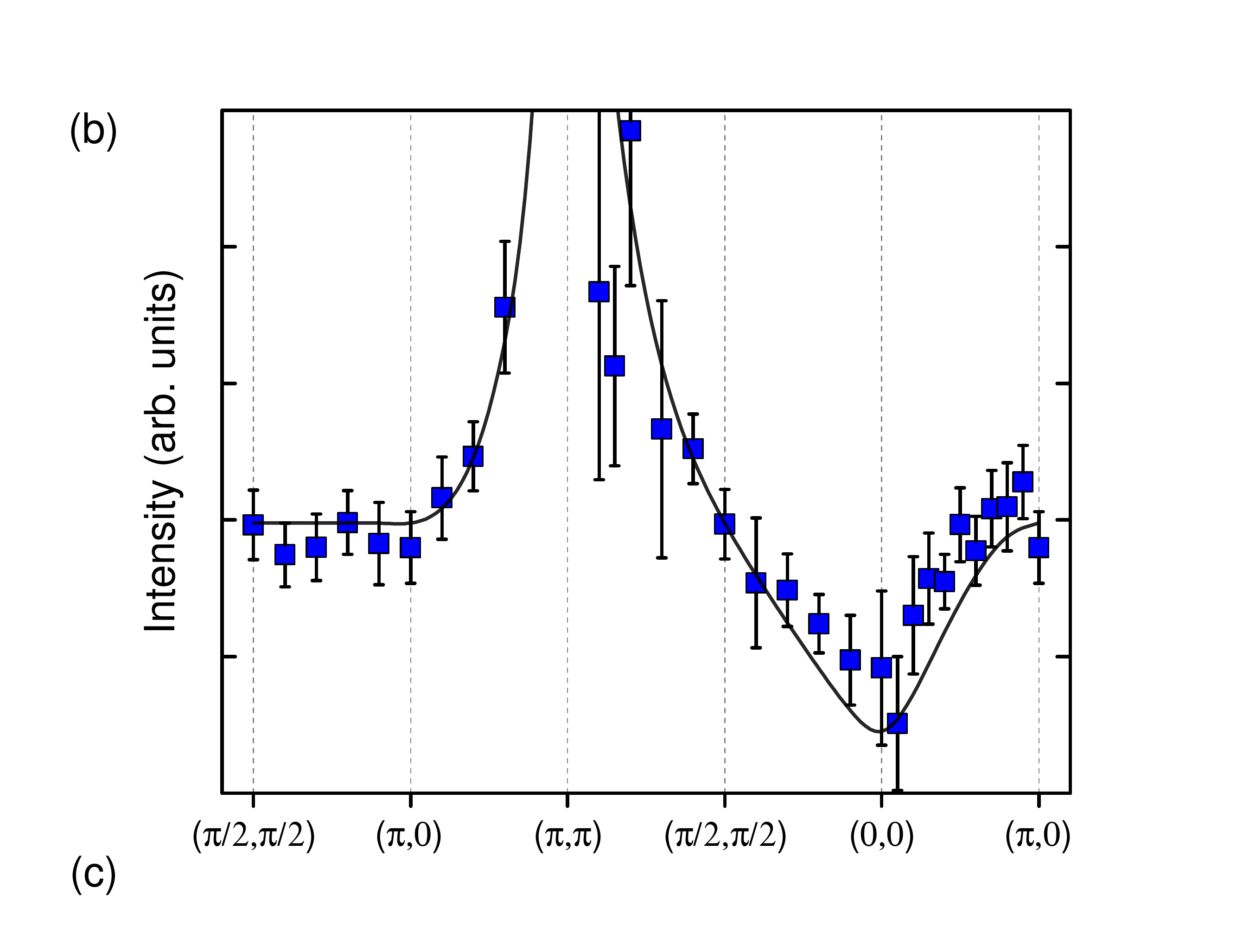}
	\includegraphics[width=0.9\columnwidth]{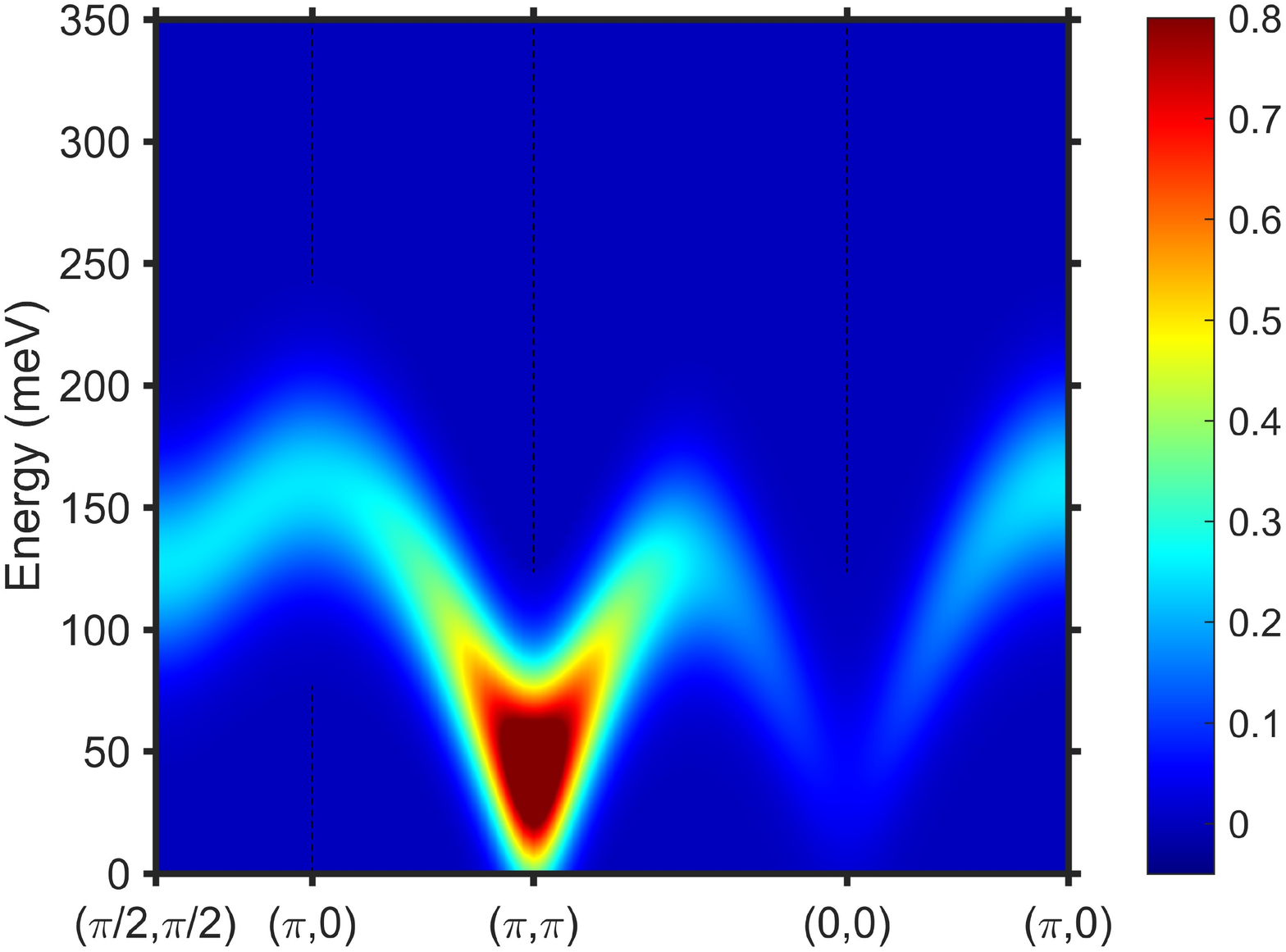}
	\caption{\label{FigFits} Measured (circles) and calculated (line) dispersion along high symmetry directions. (b) Measured (squares) and calculated (line) intensity (c) Calculated map of the dispersion and intensity.}
\end{figure}

The final point is the most striking and a departure from RIXS measurements of doping on the Sr site that shows, as with the undoped Sr$_2$IrO$_4$ case, no spin gap in the RIXS spectra \cite{PhysRevB.93.241102,PhysRevLett.117.107001} and is consistent with recent measurements on higher Ru doping \cite{Ruarxiv}. The previous lack of a measured spin-gap is somewhat surprising given the expected presence of anisotropic magnetic interactions in a strongly SOC system with large orbital component. The best RIXS resolution of $\sim$30meV, however, precluded the measurement of any small gaps but evidence of such a spin-gap was reported from electron spin resonance of 0.8 meV \cite{PhysRevB.89.180401}. This small gap, contrasting with the 90 meV giant magnon-gap of the next member of the Ruddleson-Popper series Sr$_3$Ir$_2$O$_7$ \cite{PhysRevLett.109.157402}, was explained in Ref.~\onlinecite{PhysRevB.89.180401} for Sr$_2$IrO$_4$ as being due to the quasi-degeneracy of the large antisymmetric DM anisotropy that leads to the observed net FM moment and the large symmetric anisotropies, such as single-ion and exchange anisotropy, resulting in a canceling out of in-plane and out-of-plane anisotropies and a suppressed spin-gap. Therefore following this reasoning in Sr$_2$Ir$_{0.8}$Ru$_{0.2}$O$_4$ that has $c$-axis aligned pseudospins this quasi-degeneracy is broken and a large gap can emerge. Considering points (i)-(iii) reveals an anisotropic alteration of the dispersion compared to the undoped case and, with the smaller dispersion from  ($\pi,0$) $\rightarrow$  ($\pi/2,\pi/2$) behavior that qualitatively appears closer to the cuprate case. 

To gain a quantitative understanding of the low energy excitation and underlying Hamiltonian we extracted the energy and intensity of the excitation by fitting to a Lorentzian with a sloping background and modeled the resulting dispersion within linear spin wave theory \cite{spinW}. As a natural starting point we follow the approach for Sr$_2$IrO$_4$ by using an isospin-1/2 Heisenberg model on a square lattice with $J$,$J'$,$J''$ for the first, second and third nearest neighbor exchange interactions in the basal plane. For the undoped case these values correspond to $J=60$ meV, $J'=-20$ meV and $J''=15$ meV. The undoped case did not have an observable gap in the RIXS measurements however we account for this spin-gap here by introducing a single-ion term $\mathbf{\Delta}=D\sum(S_z ^2)_i$. We note alternative terms such as Dzyaloshinskii-Moriya (DM) interaction and anisotropic exchange interaction likely contribute to this $\mathbf{\Delta}$ parameter, however we do not pursue the relative contribution of different terms further. 

Allowing the exchange interactions to refine to the measured dispersion consistently resulted in $J''$ converging towards zero, within error. The dominant effect of $J''$ on the dispersion is to decrease (positive $J''$) or increase (negative $J''$) the energy at ($\pi/2,\pi/2$) relative to ($\pi,0$) that, when non-zero, is inconsistent with the extracted dispersion data. Considering the crystal structure of  Sr$_2$Ir$_{0.8}$Ru$_{0.2}$O$_4$ one in five Ir atoms are disrupted by Ru substitution. Since a good fit for the dispersion is found without $J''$ then this indicates that longer range exchange interactions beyond nearest and next nearest neighbors are disrupted. This is distinct from the case of doping on the Sr site of Sr$_2$IrO$_4$  where $J$,$J'$,$J''$ are still needed to describe the dispersion \cite{PhysRevB.93.241102,PhysRevLett.117.107001}. Considering  $J$, $J'$ and $\Delta$ as a minimal model gives good agreement to the measured dispersion, as shown in  Fig.~\ref{FigFits}(a) with values of $J=42$ meV, $J'=-16$ meV, $J''=0$ meV and $D$=9 meV. We based our fitting on the extracted energies rather than the observed intensity due to the incomplete understanding of RIXS scattered cross sections and interferences from elastic line. However, when using the values for exchange interactions to calculate the intensity we find reasonable agreement, as shown in Fig.~\ref{FigFits}(b). Taking this further we show the purely magnetic scattering cross section from our spin-wave calculations in Fig.~\ref{FigFits}(c) that show qualitative agreement with the RIXS map we obtained once subtracting the elastic line in Fig.~\ref{Figscans}(f).

Collectively our results reveal the introduction of strong anisotropy that can be explained by the breaking of the quasi-degeneracy found in Sr$_2$IrO$_4$ that artificially suppresses the observation of a spin-gap. Given the strongly SOC dominant behavior previously observed in Sr$_2$Ir$_{0.8}$Ru$_{0.2}$O$_4$ then the measurement of such a large spin-gap provides further evidence of the importance of SOC in the magnetism, despite the dilution with Ru$^{4+}$ ions. Within this anisotropic picture, coupled with the disordered effect of doping, may suggest a departure from a purely Heisenberg model. However, we were able to apply such a model to describe the excitations in a similarly robust methodology to the parent compound. The dispersion is altered, however, from Sr$_2$IrO$_4$. The principle distinction can be explained as the Ru ions breaking the $J''_{\rm Ir-Ir}$ third nearest interactions. This behavior is in-line with the percolative picture where the dopant ions disrupt the magnetic ordering rather than interact directly with the Ir ions.

In conclusion, the RIXS spectra for Sr$_2$Ir$_{0.8}$Ru$_{0.2}$O$_4$ was measured out to 1 eV energy loss. The signature of the ${J}_{\mathrm{eff}}\mathbf{=}1/2$ state was observed as a 650 meV momentum dependent excitation. A lower energy excitation was measured with high resolution ($\Delta$E=35 meV) along high symmetry directions in the magnetic Brillouin zone. The results were well described within a 2D Heisenberg minimal model of nearest and next nearest interactions. A large spin-gap of 40 meV was observed that is consistent with the magnetic structure in Sr$_2$Ir$_{0.8}$Ru$_{0.2}$O$_4$ and a large orbital moment on the Ir ion. Modeling of the dispersion for the hole doped case here provides a contrast to the electron doped RIXS studies with alternative behavior that will guide future studies aimed at uncovering bulk superconductivity in doped Sr$_2$IrO$_4$.

\begin{acknowledgments}
The research at ORNL was sponsored by the Scientific User Facilities Division, Office of Basic Energy Sciences, U.S. Department of Energy. Use of the Advanced Photon Source, an Office of Science User Facility operated for the U.S. DOE Office of Science by Argonne National Laboratory, was supported by the U.S. DOE under Contract No. DE-AC02-06CH11357.
\end{acknowledgments}


%

\end{document}